\documentclass[aps,prb,preprint]{revtex4}
\usepackage{graphicx}

\begin{document}

\title{First-order Raman spectra of double perovskites
AB$'_{1/2}$B$''_{1/2}$O$_3$}

\author{S. A. Prosandeev$^{1,2}$, U. Waghmare$^{3}$, I. Levin$^{1}$, and J.
Maslar$^{1}$}
 \affiliation{$^1$National Institute of Standards and
Technology, Gaithersburg, Maryland 20899-8520\\
$^2$Physics Department, Rostov State University, 5 Zorge St.,
344090 Rostov on Don, Russia \\
$^3$Theoretical Sciences Unit J Nehru Centre for Advanced
Scientific Research, Jakkur PO, Bangalore 560 064, INDIA}

\date{\today}

\begin{abstract}
First principles computations of Raman intensities were performed
for perovskite-family compound CaAl$_{1/2}$Nb$_{1/2}$O$_3$~ (CAN).
This compound features 1:1 (NaCl-type) ordering of Al and Nb
superimposed onto the $b^-b^-c+$~ octahedral tilting.  Raman
tensor for CAN was computed using the package for first-principles
computations ABINIT (URL \underline {http://www.abinit.org}).
Computations performed for both untilted cubic ($Fm\overline{3}m$)
and tilted monoclinic ($P2_1/n$) CAN structures showed that the
strongest Raman lines are associated with the ordering of Al and
Nb.  The computed spectrum agreed qualitatively with the
experimental data measured on powder (CAN is available in
polycrystalline form only).  The effect of cation disorder on the
Raman intensities was considered using phenomenological theory of
light scattering in the vicinity of a phase transition. We suggest
that, for certain modes, the corresponding Raman intensities
depend primarily on the average long range order while, for other
modes, the intensities are determined by fluctuations of the order
parameter.
\end{abstract}

\maketitle

\section{Introduction}

Electronic properties of double perovskites
AB$'_{1/2}$B$''_{1/2}$O$_3$~ are strongly influenced by the degree
of B-cation ordering. \cite{Smolenski,Cross,Mitchell} Raman
spectroscopy is an attractive technique for probing B-site
ordering in perovskites. \cite{Loudon} Raman spectra proved to be
sensitive even to small nanoscale ordered regions. For example,
Raman spectra of largely disordered PbSc$_{1/2}$Ta$_{1/2}$O$_3$~
still feature well detectable lines attributed to the existence of
nanoscale ordered regions. \cite{Siny}

The quantitative theoretical description for Raman spectra of
disordered or partially ordered double perovskites has not been
developed. Furthemore, no first-principles computations of Raman
intensities in completely ordered double perovskites have been
reported. This lack of theoretical analyses of Raman intensities
impedes development of Raman spectroscopy into a quantitative tool
for probing B-cation ordering in perovskites.

Raman effect refers to a light scattering by normal vibrations
having appropriate symmetry. Landau and Placzek \cite{Placzek}
have shown that Raman intensities can be expressed over the
phonon-mediated change of high-frequency (electronic) dielectric
permittivity $\varepsilon(\omega_L)$~ where $\omega_L$~ is the
incident light frequency of a laser. The theory commonly exploits
this relation to evaluate the effect of a specific vibration on
Raman spectra. This approach requires complex computations of the
high-frequency (electronic) dielectric permittivity tensor and its
weak variations caused by ionic vibrations. First-principles
computations of Raman spectra have been limited to several
semiconductors (Si, Ge, SiO$_{2}$ etc.) and rare-earth
transition-metal borocarbides.
\cite{Baroni,Windl,Umari01,Umari00,Ravindran})

The present study reports the first attempt to compute Raman
intensities for double perovskites from first principles. The
paper is structurally as follows. Section 2 describes the general
theory for the computation of Raman intensities. Section 3
considers ordered double perovskites AB$'_{1 / 2}$B$''_{1 /
2}$O$_{3}$~ and includes three parts: the general theory of Raman
spectra for cubic double perovskites (Section 3.1), results of the
computations for the untilted cubic CaAl$_{1 / 2}$Nb$_{1 /
2}$O$_{3}$ (CAN) (Section 3.2), and results of the computations
for the tilted monoclinic CAN (Section 3.3). The effect of
B-cation disorder on the Raman spectra of double perovskites will
be described in Section 4. Section 5 considers the effect of
phonon dispersion on the Raman line profile. Finally, Section 6
summarizes the results obtained.

\section{Theory}

Born an Huang \cite{Born} derived an expression, which reduces the
computation of integral Raman intensities to the evaluation of the
imaginary part of linear (Raman) susceptibility (we consider the
part of this intensity connected with the mode $i$, polarization
directed along $\alpha $, and the field along $\beta )$:

\begin{equation}
\label{eq1} I_{i\alpha \beta } = \kappa R_{i\alpha \beta }^2
\int\limits_{ - \infty }^\infty {\left[ {n\left( \omega \right) +
1} \right]Im\chi _{i\alpha \beta } \left( \omega \right)} d\omega
\end{equation}
where $\kappa = 2\hbar \left(\omega _L - \omega _i \right)^4 /
\left ( \pi c^4\right )$, $\omega _L $ is the laser frequency,
$R_{i\alpha \beta }$~ is Raman tensor, $n( \omega ) + 1 = \left[ {1 -
\exp \left( { - \hbar \omega _i / k_B T} \right)} \right]^{ - 1}$is
Bose occupation number, $\omega _i $ is the frequency of mode $i$,
$k_B $ is Boltzmann constant, $\hbar $ is Planck constant, $T$
is temperature, and $\chi _{i\alpha \beta } $ is the contribution of
mode $i$ to dielectric susceptibility. Kramers-Kronig
transformation yields\cite{Petzelt}:

\begin{equation}
\label{eq2}
I_{i\alpha \beta } = \frac{2\hbar \left( {\omega _L - \omega _i }
\right)^4}{c^4m_i \omega _i }\left[ {n\left( {\omega _i } \right) + 1}
\right]\left( {R_{i\alpha \beta } } \right)^2
\end{equation}
Thus, the relative intensity of the Raman active modes is
determined by the fundamental characteristics of the lattice
dynamics: phonon frequency, mass, and Raman tensor.

The Raman tensor is a derivative of the electronic part of the
dielectric permittivity tensor $\varepsilon _{\alpha \beta}$~ at a
laser frequency with respect to the $i$-th normal mode
displacement:

\begin{equation}
\label{eq3} R_{i\alpha \beta } = \frac{\partial \varepsilon
_{\alpha \beta }(\omega_L)}{\partial Q_i }
\end{equation}

The first-order Raman scattering can be described using the
expansion of polarization

\begin{eqnarray}
\label{eq4} &&P_\alpha = P_{\alpha 0} + \varepsilon _0
\sum\limits_i {R_{i\alpha \beta } E_\beta Q_i } = \nonumber \\ &&
= P_{\alpha 0} + \varepsilon _0 \sum\limits_{n\gamma i} {m_i^{1 /
2} R_{i\alpha \beta ,n\gamma } E_\beta e_{in\gamma } m_n^{ - 1 /
2} Q_i }
\end{eqnarray}
where $e_{m\gamma } $ is the eigenvector of a dynamical matrix,
$m_{n}$ is the mass of the atom $n$, and $Q_{i}$ is the normal
mode displacement magnitude. From Eq. (\ref{eq4}):

\begin{equation}
\label{eq5}
R_{i\alpha \beta } = m_i^{1 / 2} \sum\limits_{n\gamma } {R_{\alpha \beta
,n\gamma } e_{in\gamma } m_{in}^{ - 1 / 2} }
\end{equation}
Eq. (\ref{eq5}) can be used to estimate the contribution of
different ions in mode $i$ into the Raman tensor.

The integral Raman intensity can be expressed as:

\begin{equation}
\label{eq6}
I_{i\alpha \beta } = \frac{2\hbar \left( {\omega _L - \omega _i }
\right)^4}{m_0 c^4\omega _i }\left[ {n\left( {\omega _i } \right) + 1}
\right]\left( {r_{i\alpha \beta } } \right)^2
\end{equation}
with

\begin{equation}
\label{eq7}
r_{i\alpha \beta } = \left( {\frac{m_i }{m_0 }} \right)^{ - 1 / 2}R_{i\alpha
\beta }
\end{equation}
where $m_0 = 1a.m.u.$

Expression (\ref{eq7}) is applicable to single crystals. For powders, the
integral
intensity has be averaged over the possible directions of crystallites which can
be accomplished using the rotation invariants

\begin{eqnarray}
\label{eq8} G_i^{(0)} & = & \frac{1}{3}\left( r_{ixx} + r_{iyy} +
                            r_{izz} \right)^2 \nonumber \\
 G_i^{(1)} & = & \frac{1}{2} [ \left( r_{ixy} - r_{iyx}
\right)^2 + \left( r_{ixz} - r_{izx}  \right)^2 + \nonumber \\
 &+& \left(r_{izy} -  r_{iyz}  \right)^2 ] \nonumber \\
 G_i^{(2)} &=& \frac{1}{2}[ \left( r_{ixy} + r_{iyx}  \right)^2 + \left(
 r_{ixz} + r_{izx}  \right)^2 + \nonumber \\ &+& \left( r_{izy} +
 r_{iyz}  \right)^2 ] +
 + \frac{1}{3}[ \left( r_{ixx} - r_{iyy} \right)^2 +
 \nonumber \\ & +&
\left( r_{ixx} - r_{izz}  \right)^2 + \left( r_{iyy} - r_{izz}
\right)^2 ]
\end{eqnarray}
For the 90$^{0}$ and backscattering geometry, a (polarized) laser
beam, and a scattered radiation analyzer in use, the (reduced)
intensity for the powder can be expressed as
\cite{Placzek,Hayes,Hiro}:

\begin{equation}
\label{eq9}
\begin{array}{l}
 I_{\parallel}^{powder} \sim \left( {\omega _L - \omega _i } \right)^4\frac{1 +
n\left( {\omega _i } \right)}{30\omega _i }\left[ {10G_i^{(0)} +
4G_i^{(\ref{eq2})}
} \right], \\
 I_{\perp}^{powder} \sim \left( {\omega _L - \omega _i } \right)^4\frac{1 +
n\left( {\omega _i } \right)}{30\omega _i }\left[ {5G_i^{(\ref{eq1})} +
3G_i^{(\ref{eq2})} }
\right], \\
 I_{tot}^{powder} = I_{\parallel}^{powder} + I_{\perp}^{powder} \\
 \rho _i^{powder} = I_{\perp}^{powder} / I_{\parallel}^{powder} \\
 \end{array}
\end{equation}
where subscripts $\parallel$ and $\perp$ correspond to the
polarized and depolarized light, respectively.

Finally, one can plot the Raman spectrum by using the
expressions:

\begin{equation}
\label{eq10}
\begin{array}{l}
 I_{\vert \vert }^{powder} \left( \omega \right)\sim \sum\limits_i
{I_{i\vert \vert }^{powder} \frac{\Gamma _i }{\left( {\omega - \omega _i }
\right)^2 + \Gamma _i^2 }} \\
 I_ \bot ^{powder} \left( \omega \right)\sim \sum\limits_i {I_{i \bot
}^{powder} \frac{\Gamma _i }{\left( {\omega - \omega _i } \right)^2 + \Gamma
_i^2 }} \\
 I_t^{powder} \left( \omega \right)\sim \sum\limits_i {I_{it}^{powder}
\frac{\Gamma _i }{\left( {\omega - \omega _i } \right)^2 + \Gamma _i^2 }} \\
 \end{array}
\end{equation}
where $\Gamma _{i}$ is the damping constant for the $i$-th mode.
The computation of this constant should be done separately. Here
we concentrate on the computation of integral Raman
intensities only.

\section{Raman spectra of ordered AB$'_{1 / 2}$B$''_{1 /
2}$O$_{3}$ perovskites}

\subsection{Results of first-principles computations for cubic
CaAl$_{0.5}$Nb$_{0.5}$O$_{3}$}

CaAl$_{0.5}$Nb$_{0.5}$O$_{3}$ (CAN) crystallizes with a perovskite-like
structure and exhibits 1:1 ordering of Al and Nb superimposed onto the
 $b^-b^-c^+$~ tilting of oxygen octahedra.\cite{Cockayne} This
combination of cation ordering and octahedral tilting yields
monoclinic $P2_1 / n$symmetry with 20 atoms per unit
cell.\cite{Levin} In the absence of tilting, the structure would
adopt a cubic $Fm\overline{3}m$~ symmetry yielding $A_{1g} + E_g +
2F_{2g} $ Raman-active modes (see Fig. \ref{Fig1} and Appendix).
First-principles computations of vibration frequencies, dynamical
charges, and infra-red spectra for the cubic and tilted monoclinic
structures of CAN have been reported.\cite{Cockayne} The present
study focuses on the computation of Raman intensities with the
help of a package for \textit{ab-initio} computations, ABINIT,
which  is a result of joint project between the Universit\'{e}
Catholique de Louvain, Corning Incorporated, and other
contributors. \cite{Gonze02} Ingredients needed for the
calculation of responses to atomic displacements and homogeneous
electric fields have been already described in the literature.
\cite{Gonze97_1,Gonze97_2} We do not compute damping constants in
the present study, only integral intensities. A damping constant
$\Gamma _{i}$ = 10 cm$^{ - 1}$ was used for all modes.

First, we performed the first-principles computations for the
10-ion unit cell of CAN having symmetry $Fm\overline{3}m$~ (no
tilting). We used 8x8x8 Monkhorst-pack grid, ECUT=45 Ha and tolvrs
= tolwfr = 10$^{ - 18}$. We computed high frequency (electronic)
dielectric tensor for the 5 different normal mode displacements.
Then, we interpolated the results with a polynomial and extracted
the linear dependence of the high-frequency (electronic)
dielectric permittivity on the normal mode displacement. The
results of the computation of the Raman tensors are summarized in
Table 1. The $A_{1g}$ mode has the largest Raman tensor. Table 2

presents the coefficients needed for the computation of the
average Raman tensor for the $F_{2g}$ modes. Results of the
computation of Raman intensities are presented in Table 3 and Fig.
\ref{Fig2}.

\subsection{Raman spectrum of tilted CAN structure (P2$_{1}$/n)}

For the tilted CAN ($P$2$_{1}$/$n)$, we used the eigenfrequencies
and eigenvectors which are similar to those obtained by Cockayne.
\cite{Cockayne} The lattice constants in these computations were
taken from the experiment, but the distortion and the equilibrium
atomic coordinates were determined selfconsistently under the
constraint of a fixed volume. The obtained structure contains
frozen tilting of the $b^ - b^ - c^ + $ type in the Glazer
notations, and the equilibrium structure is monoclinic $P2_1/n$
with the unit cell 4 times larger than the primitive perovskite
cell. By using ABINIT, we computed high-frequency (electronic)
dielectric tensor, $\varepsilon _{\alpha \beta }(\omega_L) $, for
each of the displacements $Q$. In all cases, we used the same high
precision and 8x8x8 Monkhorst-pack grid as in the computation for
the cubic CAN. To facilitate the comparison with experiment, we
used the damping constants of 8 cm$^{ - 1}$ and 2 cm$^{ - 1}$~ at
frequencies above and below 400 cm$^{ - 1}$, respectively.

Fig. \ref{Fig3} presents the experimental (a) and the calculated
total (b), polarized (c), and depolarized (d) Raman spectra for a
ceramic CAN (the integral intensities for all lines are shown in
Table 5).

\section{Disordered AB$'_{1/2}$B$''_{1/2}$O$_{3}$ perovskites}

Raman spectra of double perovskites reportedly are sensitive to
the degree of cation order, \textit{$\eta$}.\cite{Siny} The 1:1
order parameter for the $Pm\bar {3}m \to Fm\bar {3}m$ transition
is a scalar and transforms according to the $R_{1}^{ + }$ representation
of the parent $Pm\bar {3}m$ space group. A useful insight into the
dependence of Raman intensities on \textit{$\eta $} can be
obtained using phenomenological theories of light scattering in
the vicinity of a phase transition.\cite{Petzelt,Ginzburg}
We start with the expression (\ref{eq7})and expand the Raman
tensor into a series with respect to the order parameter:

\begin{equation}
\label{eq31} r_{i\alpha \beta } ({\rm{\bf r}}) = r_{i\alpha \beta
,1} \eta({\rm{\bf r}}) + ...
\end{equation}
The change of dielectric permittivity $\varepsilon({\rm{\bf r}})$~
is proportional to the product of $\eta({\rm{\bf r}})$~ and
displacement $Q({\rm{\bf r}})$. The Fourier transform of this
product can be expressed as:

\begin{equation}
\Delta \varepsilon_q \sim \sum_{\rm{\bf k}}{ \eta_{{\rm {\bf
k}}+{\rm {\bf q}}} Q_{-{\rm{\bf k}}}}
\end{equation}
Finally, the intensity of the light scattering is proportional to
the statistical average of the square of $\left | \Delta
\varepsilon_q \right |$:

\begin{equation}
I \sim \sum_{{\rm{\bf k k}}'}{ \left <\eta_{{\rm {\bf k}}+{\rm
{\bf q}}} Q_{-{\rm {\bf k}}}\eta_{-{\rm {\bf k}}'-{\rm {\bf q}}}
Q_{{\rm {\bf k}}'}\right
> } \label{I_General}
\end{equation}
According to this expression, phonons in the entire Brillouin zone
are excited if $|\eta| < 1$.

Recent experimental studies of $\eta$-dependence of Raman
intensities in CAN have shown that the intensities corresponding
to both $F_{2g}$- and $E_{g}$-related modes vary approximately as
square of the order parameter.\cite{APL} In contrast, the
intensity of the $A_{1g}$ line exhibited weak dependence on $\eta
$. These observation can be understood by considering two
particular cases of the general expression (\ref{I_General}).

First, we consider the case when $Q$~ does not correlate with
$\eta$:

\begin{equation}
I^{(1)} \sim \sum_{{\rm{\bf k}}} \left < \eta_{{\rm {\bf k}}+{\rm
{\bf q}}} \eta_{-{\rm {\bf k}}-{\rm {\bf q}}}\right> \left <
Q_{\rm{\bf k}}Q_{-{\rm{\bf k}}} \right > \label{eta2}
\end{equation}
Now, we present $\eta({\rm{\bf r}})$ as the sum of the average
order parameter, $\eta_e$~ and its fluctuation $\Delta
\eta({\rm{\bf r}})$:

\begin{eqnarray}
&& I^{(1)}  \sim \eta_e^2 \left < Q_{\rm{\bf q}}Q_{-{\rm{\bf q}}}
\right
> + \nonumber \\ && + \sum_{{\rm{\bf k}}} \left <\Delta \eta_{{\rm {\bf k}}+{\rm
{\bf
q}}} \Delta \eta_{-{\rm {\bf k}}-{\rm {\bf q}}}\right> \left <
Q_{\rm{\bf k}}Q_{-{\rm{\bf k}}} \right > \label{first}
\end{eqnarray}
The second term is finite only in the vicinity of the phase
transition temperature, and it is finite even above $T_c$~ (the
${\rm{\bf k}}=0$~ contribution behaves\cite{Ginzburg} as $
|\tau|^{-1/2}$). The first term behaves as:

\begin{equation}
\eta_e^2 \sim \tau = \frac{T_c - T_a }{T_a } \label{square}
\end{equation}
where $T_{c}$~ is the order -- disorder phase transition
temperature, $T_{a}$~ the annealing temperature (the equation can
be applied to the samples quenched from $T_{a})$. This type of the
dependence was experimentally observed for the $F_{2g}$~ and
$E_g$~ related modes in CAN. \cite{APL}

Now we consider the case when $Q$~ and $\eta$ are strongly
correlated and the displacements are modulated by the order
parameter:

\begin{equation}
\label{eq33} Q( {\rm {\bf r}}) \sim \eta \left( {\rm {\bf r}}
\right)
\end{equation}
In particular, the displacements change sign at the antiphase
domain boundaries following the change in the sign of the order
parameter. so that changing the sign of the order parameter has no
effect on the Raman intensity. Even at the antiphase domain wall,
where $\eta$~ varies linearly from --1 to 1, one gets a finite
contribution to the intensity. Relation (\ref{eq33}) is realized
only for those modes which transform similarly to the order
parameter. In the symmetry group, $Fm\overline{3}m$, only the
$A_{1g}$~ mode transforms similarly to $R1^+$ mode in
$Pm\overline{3}m$.

In this case, (\ref{I_General}) takes the form:

\begin{equation}
\label{eq36}
\begin{array}{l}
 I_{it}^{(\ref{eq2})} = C_i \eta _e^2 \left\langle {\Delta \eta _{\rm {\bf q}}
\Delta \eta _{ - {\rm {\bf q}}} } \right\rangle + D_i \eta _e^4 + \\

 + E_i \sum\limits_{{\rm {\bf \kappa }},{\rm {\bf \upsilon }}} {\left\langle
{\Delta \eta _{{\rm {\bf q}} + {\rm {\bf \kappa }}} \Delta \eta _{
- {\rm {\bf \kappa }}} \Delta \eta _{ - {\rm {\bf q}} + {\rm {\bf
\upsilon }}}
\Delta \eta _{ - {\rm {\bf \upsilon }}} } \right\rangle } + \\
 + F_i \eta _e \sum\limits_{\rm {\bf \kappa }} {\left\langle {\Delta \eta
_{{\rm {\bf q}} + {\rm {\bf \kappa }}} \Delta \eta _{ - {\rm {\bf
\kappa }}}
\Delta \eta _{ - {\rm {\bf q}}} } \right\rangle } ... \\
 \end{array}
\end{equation}
where \cite{Ginzburg}

\begin{equation}
\label{eq37}
 \left\langle {\left| {\Delta \eta _{\rm {\bf q}} } \right|^2} \right\rangle
\sim \tau^{-1} \\
\end{equation}
The first term is temperature independent below $T_{c}$ and should
vanish above $T_{c}$. The second term is negligible in the
vicinity of a phase transition (it varies as $\tau^2$), and the
last two terms are spread over phonon bands (they are finite above
$T_c$). The third term depends\cite{Ginzburg} on $\tau$~ as
$|\tau|^{-1/2}$). Experimental observations \cite{APL} suggest
that, the main contribution to light scattering for the $A_{1g}$~
mode comes from the fist term in (39) and is temperature
independent.

Thus, we conclude that the $F_{2g}$~ and $E_g$~ modes obey
dependence (\ref{square}) whereas $A_{1g}$~ mode shows mostly
constant intensity below $T_a=T_c$. This difference stems from the
fact that the $A_{1g}$~ mode transforms in the same way as the
chemical order parameter while the $F_{2g}$~ and $E_g$~ modes
transform differently.

The $F_{2g}$ and $E_{g}$ modes interact with strains, which can be
constructed from the components of the gradient of the order
parameter. This interaction mixes the polarized and depolarized
spectra.

In the distorted (tilted) structure of CAN, there are Raman lines
originating presumably from tilting but not from the chemical
ordering. A ferroellastic order parameter should supplement
$\eta$, in this case.\cite{Petzelt}

\section{Phonon band width}

Disorder can cause excitation of the $q$-points in the entire
Brillouin zone.\cite{Yacoby,Richter,Kitajima,Rogacheva} The
contribution of this effect to the width of the Raman line is
limited by the width of the corresponding phonon band. We have
performed first-principles computation of phonon frequencies for a
80-ion cubic supercell of CAN (Viena DFT package VASP
\cite{Krese1,Krese2} was used) and compared the results with those
obtained for the tilted 20-ion supercell \cite{Cockayne} (Fig.
\ref{Fig4}). According to our results, the upper branch of the
phonon bands (which includes the point corresponding to the
$A_{1g}$~ mode) has the width of about 70 cm$^{ - 1}$, which is
comparable to the width of the $A_{1g}$~ Raman line detected
experimentally (about 40 cm$^{-1}$). Such narrow width increased
only slightly with temperatures approaching $T_c$, which can be
attributed to relatively narrow width of the corresponding phonon
band. This relatively narrow width of the high frequency phonon
bands in CAN can be caused by the band folding due to a doubling
of the unit cell upon ordering.

\section{Summary}

We have performed \textit{ab initio} computations of Raman
intensities for the AB$'_{1 / 2}$B$''_{1 / 2}$O$_{3}$ perovskites.
The computation of the Raman intensities in cubic CAN revealed
that all Raman-active lines, A$_{1g}$+E$_{g}$+2F$_{2g}$, exhibit
large relative intensities in single crystals.
The low-frequency part of the Raman spectrum computed for the
untilted cubic CAN structure differs significantly from the
corresponding part of the Raman spectrum for the tilted monoclinic
CAN. The intense lines in the cubic structure at 101 cm$^{ - 1}$
shift to higher frequencies, split (188 cm$^{ - 1}$, 242 cm$^{ -
1}$, and 211 cm$^{ - 1})$,\cite{Cockayne} and become relatively
weak. This is a result of the change in frequency, occupation
number, and Raman tensor for this line. The high-frequency
$F_{2g}$~ line in the cubic structure at 445 cm$^{ - 1}$ splits in
the low-symmetry structure into three lines at 446 cm$^{ - 1}$,
452 cm$^{ - 1}$ and 457 cm$^{ - 1}$.\cite{Cockayne} The relative
intensities of these lines decrease in the distorted structure
though to a lesser extent than for the low-frequency F$_{2g}$
line. E$_{g}$ line at 660 cm$^{ - 1}$ splits into two lines at 553
cm$^{ - 1}$ and 554 cm$^{ - 1}$,\cite{Cockayne} and the integral
intensity for this line decreases. A$_{1g}$ line at 910 cm$^{ -
1}$ in the cubic structure exhibits significant downshift to 797
cm$^{ - 1}$ in the distorted CAN.\cite{Cockayne} Experimentally,
this line is observed at higher frequency of 844 cm$^{ - 1}$. This
discrepancy between the experimental and computed frequencies can
be due to the overestimated degree of tilting in the DFT
computations. The computed Raman intensities agree qualitatively
with those observed experimentally. Yet, the experiment yields
relative integral intensity of the A$_{1g}$ line larger than the
computed value (mostly due to the larger width), and the computed
$E_g$ intensity is smaller in the computation in comparison with
experiment. This can be related to the
contribution from the scattering on the powder surfaces. Second
order Raman processes can contribute to the detected line
intensity as well.

On the basis of the present first-principles computation we
 conclude that: (i) In agreement with earlier speculations,
\cite{Siny} high-frequency Raman lines appear in
AB$'_{1/2}$B$''_{1/2}$O$_{3}$~ double perovskites mainly due the
ordering of B$'$~ and B$''$~ ions; (ii) The modulation of the
ionic vibrations by the order parameter in CAN can explain the
weak dependence of the $A_{1g}$~ line intensity in CAN on the
annealing temperature when approaching $T_c$; (iii) Our
first-principles computation has shown that the width of the
high-frequency phonon band is about 70 cm$^{-1}$ that can explain
the weak dependence of the high-frequency Raman lines asymmetry in
CAN on the degree of disorder.

\section*{Acknowledgement}

We are grateful to Arkady Levanyuk, Jan Petzelt, Oleksiy
Svitelskiy, Sergey Vakhrushev, Eric Cockayne, Ronald Cohen, Andrew
Rappe, Benjamin Burton and David Vanderbilt for stimulating
discussions. S.A.P appreciates grants ru.01.01.037 (``Russian
Universities''), 04-02-16103 (RFBR), and support from the NIST
MSEL's Director's Reserve Fund. We express our appreciation to
Xavier Gonze and Francois Detraux for their help with using
ABINIT.

\appendix
\section{General expressions for cubic $Fm\bar{3}m$ structure}

\begin{figure*}
\resizebox{1.0\textwidth}{!}{\includegraphics{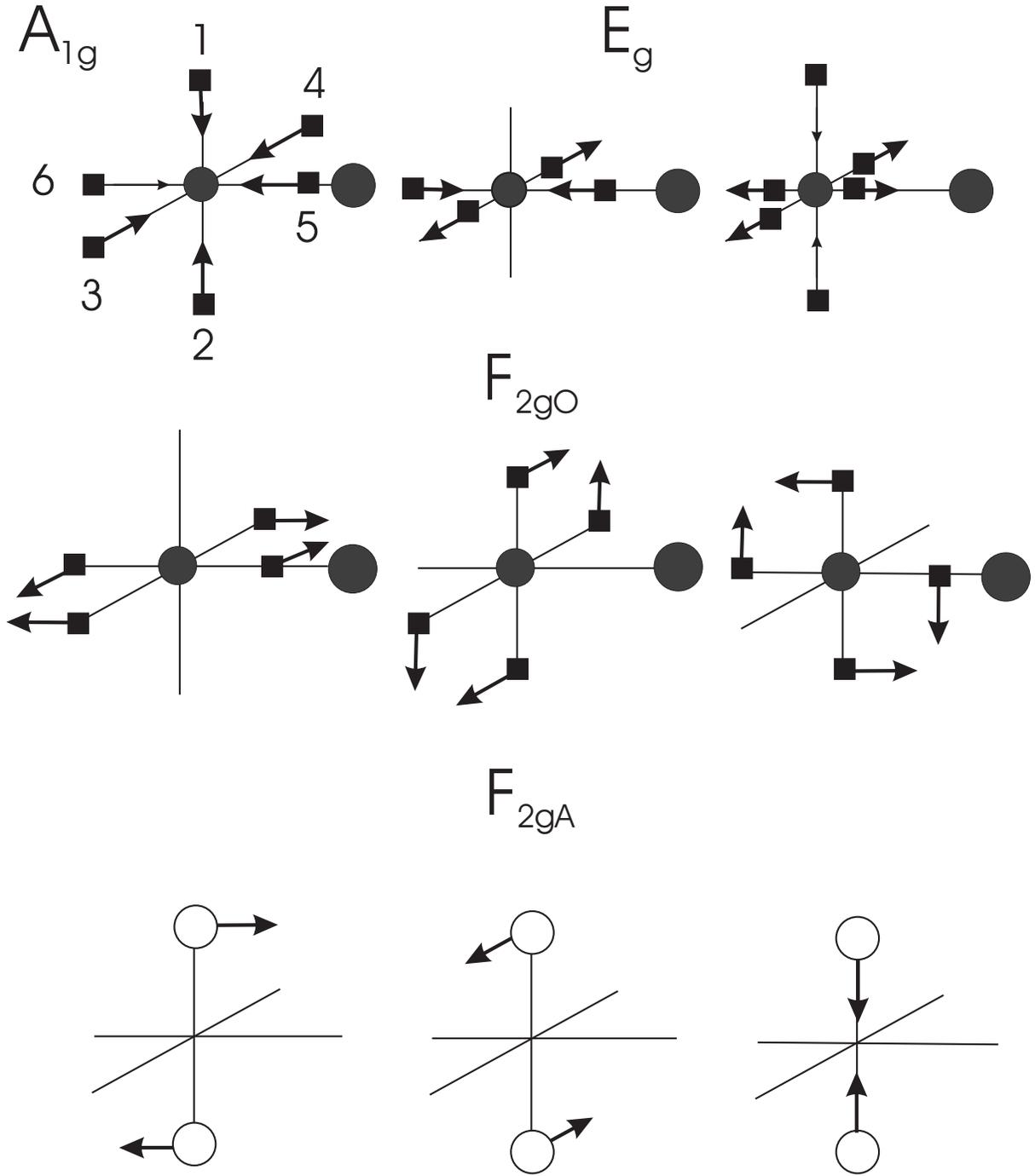}}
\caption{The Raman active modes in AB$'_{1/2}$B$''_{1/2}$O$_3$~
ordered perovskites ($Fm\overline{3}m$). Small and large filled
circles denote ions B$'$~ and B$''$, filled squares are oxygens,
and empty circles are ions A. } \label{Fig1} \end{figure*}

\begin{figure*}
\resizebox{1.0\textwidth}{!}{\includegraphics{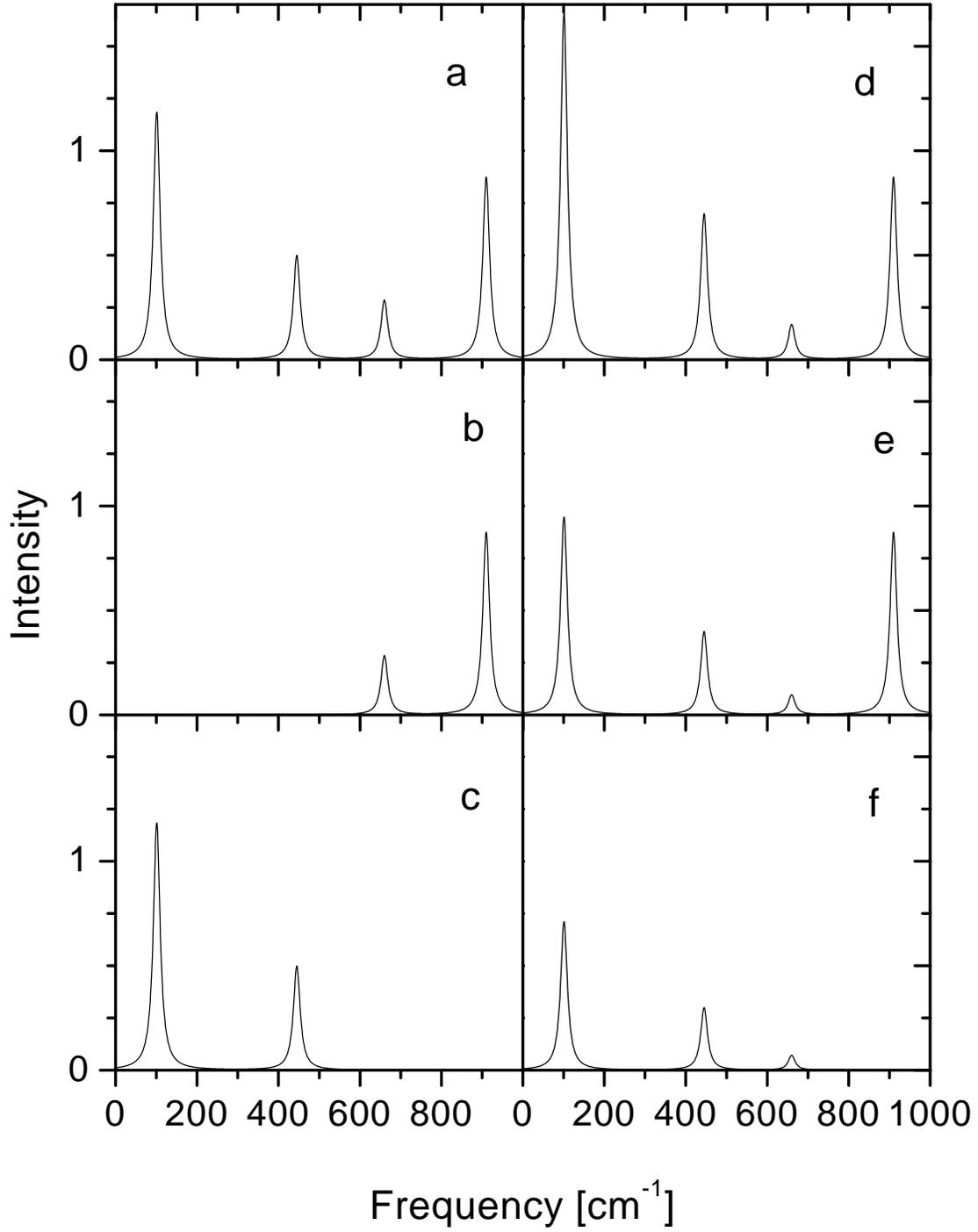}}
\caption{Computed Raman spectrum of cubic CAN: $I_t^{cryst}$~ (a),
$I_{\parallel}^{cryst}$~ (b), $I_{\perp}^{cryst}$~(c),
$I_t^{powder}$~ (d), $I_{\parallel}^{powder}$~ (e),
$I_{\perp}^{powder}$~ (f).} \label{Fig2}
\end{figure*}

\begin{figure*}
\resizebox{1.0\textwidth}{!}{\includegraphics{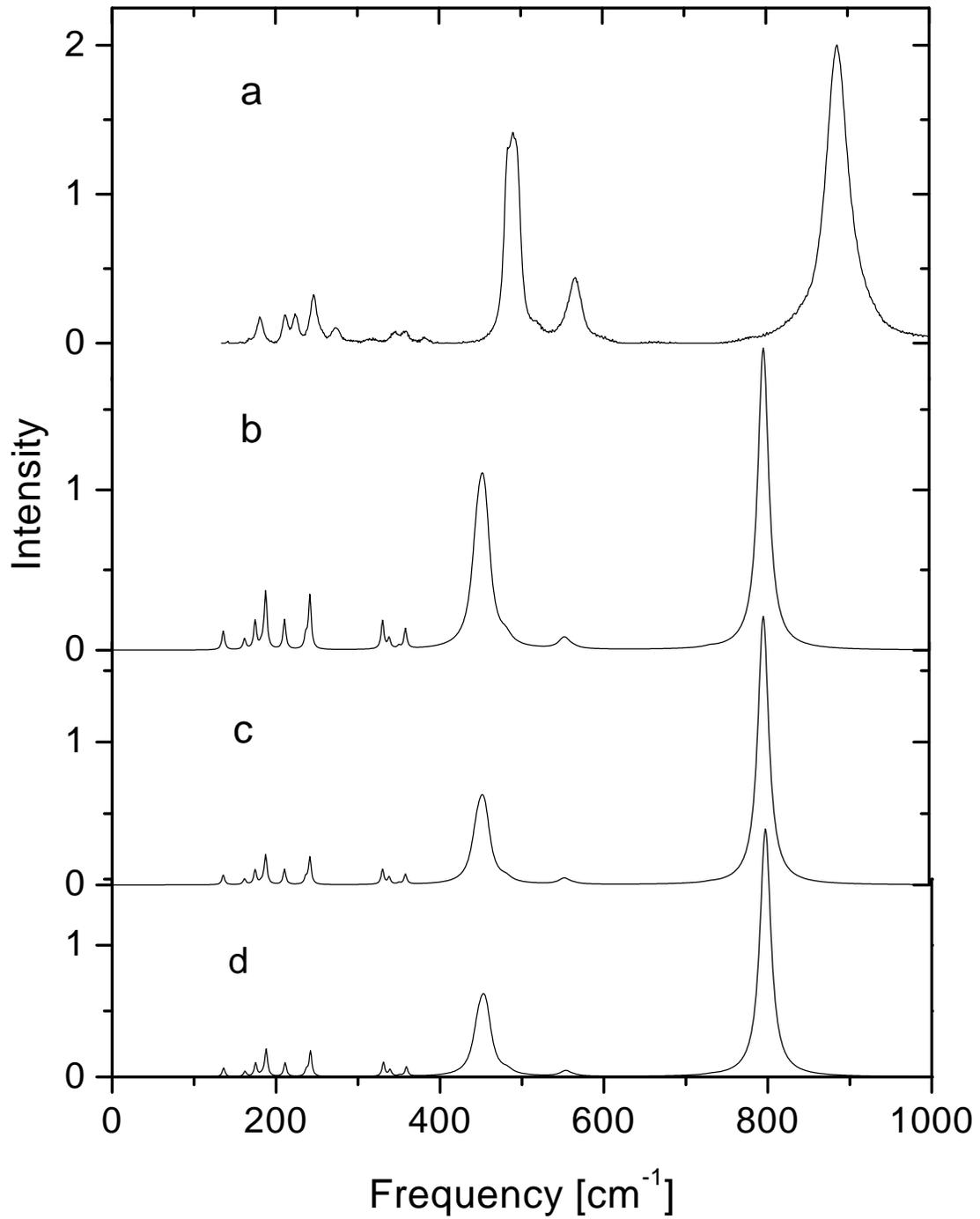}}
\caption{Computed Raman intensities for ceramic CAN ($P2_1/n$):
experiment \cite{Levin,APL} (a), theory: $I_t^{powder}$~ (b),
$I_{\parallel}^{powder}$~ (c), $I_{\perp}^{powder}$~ (d).}
\label{Fig3}
\end{figure*}

\begin{figure}
\resizebox{1.0\textwidth}{!}{\includegraphics{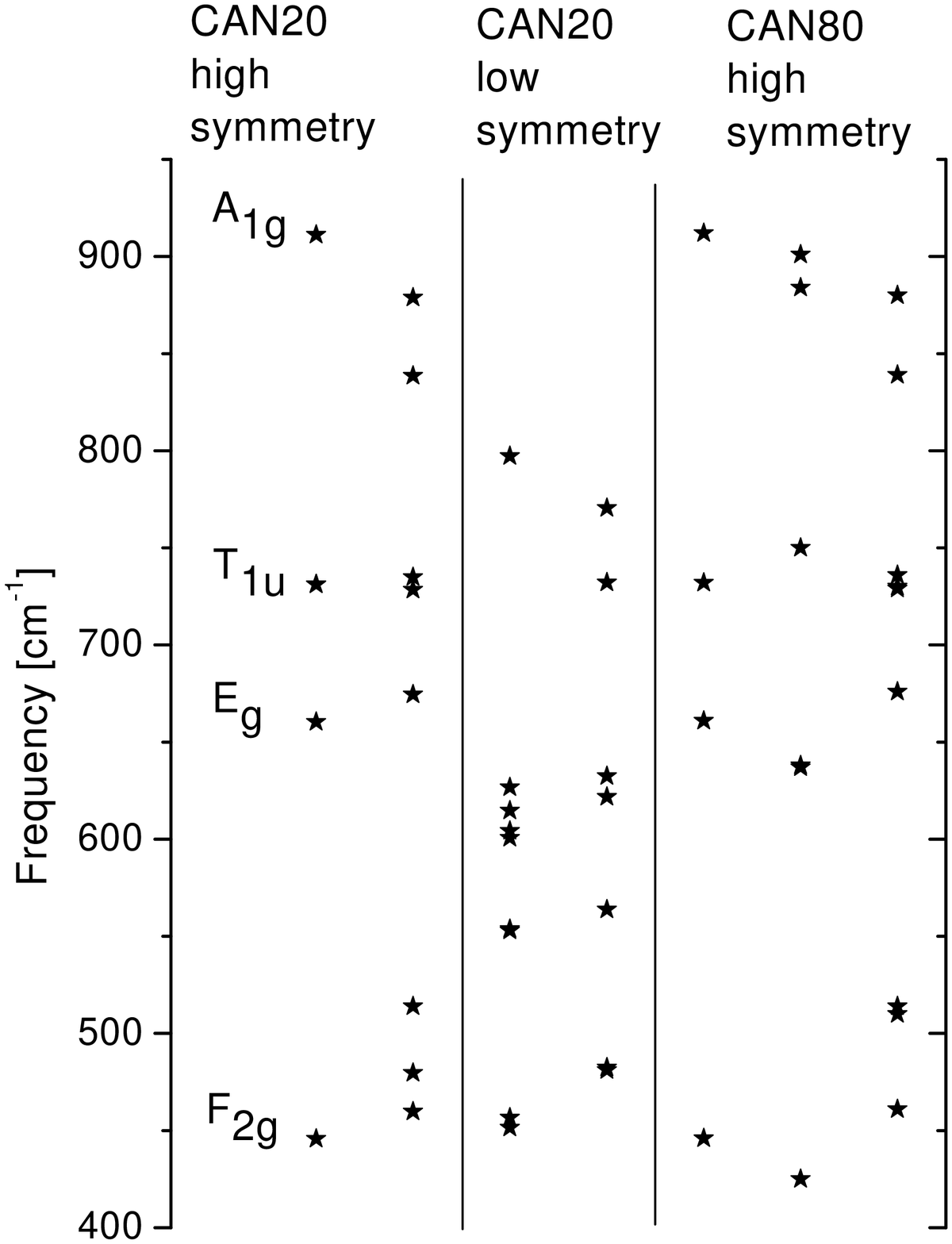}}
\caption{Results of first principles computations of phonon
frequencies in different supercells of CAN: the data for the
20-ion supercell \cite{Cockayne} are shown for comparison with our
computation for 80-ion supercell of CAN. } \label{Fig4}
\end{figure}

Cubic $Fm\bar {3}m$ ordered perovskite AB$'_{1 / 2}$B''$_{1 /
2}$O$_{3}$ exhibits Raman active modes$~ A_{1g} + E_g + 2F_{2g} $.
$A_{1g} $ is a fully symmetric breathing vibration of oxygen
octahedra (Fig. \ref{Fig1}). The Raman tensor for this mode is
diagonal,

\begin{equation}
\label{eq11} r_{A_{1g} } = \left( {{\begin{array}{*{20}c}
 a \hfill & 0 \hfill & 0 \hfill \\
 0 \hfill & a \hfill & 0 \hfill \\
 0 \hfill & 0 \hfill & a \hfill \\
\end{array} }} \right),
\end{equation}
with the diagonal element

\begin{equation}
\begin{array}{l}
\label{eq12} a = r_{A_{1g} zz} = \left ( 6m_\mathrm{O} \right
)^{-1/2} ( r_{zz,\mathrm{O}1z} - r_{zz,\mathrm{O}2z} +
r_{zz,\mathrm{O}3x} -
\\ - r_{zz,\mathrm{O}4x} + r_{zz,\mathrm{O}5y} -
r_{zz,\mathrm{O}6y}  )
\end{array}
\end{equation}
here the oxygen atoms in an octahedron are numbered from 1 to 6
(Fig. \ref{Fig1}).

Symmetry of the lattice requires that:

\begin{eqnarray}
\label{eq13}
 && r_{zz,\mathrm{O}2z} = - r_{zz,\mathrm{O}1z} = r_{xx,\mathrm{O}3x} = -
r_{xx,\mathrm{O}4x} = \nonumber
 \\ && = r_{yy,\mathrm{O}5y} = -
r_{yy,\mathrm{O}6y} \nonumber \\ &&
 r_{zz,\mathrm{O}3x} = - r_{zz,\mathrm{O}4x} = r_{zz,\mathrm{O}5y} = -
r_{zz,\mathrm{O}6y} =
 \nonumber\\
&& = r_{xx,\mathrm{O}1z} = - r_{xx,\mathrm{O}2z} =
r_{xx,\mathrm{O}5y} = -
r_{xx,\mathrm{O}6y} = \nonumber \\
&& = r_{yy,\mathrm{O}1z} = - r_{yy,\mathrm{O}2z} =
r_{yy,\mathrm{O}3x} = -
r_{yy,\mathrm{O}4x} \\
\end{eqnarray}
Hence, (\ref{eq12}) reduces to:

\begin{equation}
\label{eq14} r_{A_{1g} zz} = 2\sqrt {\frac{m_0 }{6m_\mathrm{O} }}
\left( {r_{zz,\mathrm{O}1z} + 2r_{zz,\mathrm{O}3x} } \right)
\end{equation}
where $m_{\mathrm{O}}$ is the mass of an oxygen atom. Thus, the
Raman tensor for this mode is determined by the two quantities
describing the longitudinal and transverse changes of polarization
with respect to atomic displacements.

$E_{g}$ mode corresponds to another type of breathing of oxygen
octahedra (Fig. \ref{Fig1}). This mode is doubly degenerate. The
corresponding Raman tensors can be represented as:

\begin{equation}
\label{eq15} r_{E_g (\ref{eq1})} = \left( {{\begin{array}{*{20}c}
 { - b} \hfill & 0 \hfill & 0 \hfill \\
 0 \hfill & { - b} \hfill & 0 \hfill \\
 0 \hfill & 0 \hfill & {2b} \hfill \\
\end{array} }} \right),\,\,\,\,\,\,r_{E_g (\ref{eq2})} = \left(
{{\begin{array}{*{20}c}
 c \hfill & 0 \hfill & 0 \hfill \\
 0 \hfill & { - c} \hfill & 0 \hfill \\
 0 \hfill & 0 \hfill & 0 \hfill \\
\end{array} }} \right)
\end{equation}
where

\begin{equation}
\begin{array}{l}
\label{eq16} b = \frac{1}{2}r_{E_g (\ref{eq1})zz} =
\frac{1}{4\sqrt {3m_\mathrm{O} } }( 2r_{zz,\mathrm{O}1z} -
2r_{zz,\mathrm{O}2z} - \\ - r_{zz,\mathrm{O}3x} +
r_{zz,\mathrm{O}4x} - r_{zz,\mathrm{O}5y} + r_{zz,\mathrm{O}6y} )
\\
c = r_{E_g (2)zz} = \frac{1}{2\sqrt {m_\mathrm{O} } }(
r_{zz,\mathrm{O}3x} - \\ - r_{zz,\mathrm{O}4x} -
r_{zz,\mathrm{O}5y} + r_{zz,\mathrm{O}6y}  )
\end{array}
\end{equation}
From the symmetry relations (\ref{eq13}):

\begin{equation}
\label{eq17}
\begin{array}{l}
 b = \frac{1}{\sqrt {3m_\mathrm{O} } }\left( {r_{zz,\mathrm{O}1z} -
r_{zz,\mathrm{O}3x} } \right) \\
 c = \frac{1}{\sqrt {m_\mathrm{O} } }\left( {r_{zz,\mathrm{O}1z} -
r_{zz,\mathrm{O}3x} } \right) \\
 \end{array}
\end{equation}
Therefore, the Raman tensor for the $E_{g}$ modes is controlled by
the same two constants as the Raman tensor for the $A_{1g}$ mode,
and

\begin{equation}
\label{eq18} c = \sqrt 3 b.
\end{equation}
$F_{2g}$~ modes are triply degenerate and contain contributions
from the oxygen and A-cations.  Oxygen-related $F_{2g\mathrm{O}}$
modes are presented in Fig. \ref{Fig1}. Their Raman tensors are
nondiagonal:

\begin{eqnarray}
\label{eq19}
 r_{F_{2g} \mathrm{O}(\ref{eq1})} &=& \left(
{{\begin{array}{*{20}c}
 0 \hfill & {d_\mathrm{O} } \hfill & 0 \hfill \\
 {d_\mathrm{O} } \hfill & 0 \hfill & 0 \hfill \\
 0 \hfill & 0 \hfill & 0 \hfill \\
\end{array} }} \right) \nonumber \\ r_{F_{2g} \mathrm{O}(\ref{eq2})} &=& \left(
{{\begin{array}{*{20}c}
 0 \hfill & 0 \hfill & {d_\mathrm{O} } \hfill \\
 0 \hfill & 0 \hfill & 0 \hfill \\
 {d_\mathrm{O} } \hfill & 0 \hfill & 0 \hfill \\
\end{array} }} \right) \nonumber \\ r_{F_{2g} \mathrm{O}(\ref{eq3})} &=& \left(
{{\begin{array}{*{20}c}
 0 \hfill & 0 \hfill & 0 \hfill \\
 0 \hfill & 0 \hfill & {d_\mathrm{O} } \hfill \\
 0 \hfill & {d_\mathrm{O} } \hfill & 0 \hfill \\
\end{array} }} \right)
\end{eqnarray}
where

\begin{equation}
\label{eq20} d_\mathrm{O} = r_{F_{2g} \mathrm{O}(\ref{eq1}),xy} =
\frac{1}{2\sqrt {m_\mathrm{O} } }\left( {r_{xy,\mathrm{O}3x} -
r_{xy,\mathrm{O}4x} } \right)
\end{equation}
Because of the symmetry, only two ions out of six (those residing
on axis $x$) contribute to this Raman tensor. The symmetry
relations give:

\begin{equation}
\label{eq21}
\begin{array}{l}
 r_{xy,\mathrm{O}3x} = - r_{xy,\mathrm{O}4x} = r_{xz,\mathrm{O}3x} = \\ = -
r_{xz,\mathrm{O}4x} = r_{yz,\mathrm{O}5y} = -
r_{yz,\mathrm{O}6y} = \\
 = r_{yx,\mathrm{O}5y} = - r_{yx,\mathrm{O}6y} = r_{zx,\mathrm{O}1z} = \\ = -
r_{zx,\mathrm{O}2z} = r_{zy,\mathrm{O}1z} = - r_{zy,\mathrm{O}2z}
 \end{array}
\end{equation}
Thus, only one constant determines the Raman tensors for the
F$_{2g \mathrm{O}} $~ modes:

\begin{equation}
\label{eq22} d_\mathrm{O} = \frac{1}{\sqrt {m_\mathrm{O} }
}r_{xy,\mathrm{O}3x}
\end{equation}
The A-cation related ($F_{2g\mathrm{A}})$ modes (Fig. \ref{Fig1})
have the same Raman tensor as the oxygen related F$_{2g}$ modes:

\begin{eqnarray}
\label{eq23} r_{F_{2g\mathrm{A}}(\ref{eq1})} &=& \left(
{{\begin{array}{*{20}c}
 0 \hfill & {d_{\mathrm{A}} } \hfill & 0 \hfill \\
 {d_{\mathrm{A}} } \hfill & 0 \hfill & 0 \hfill \\
 0 \hfill & 0 \hfill & 0 \hfill \\
\end{array} }} \right) \nonumber \\ r_{F_{{2g} \mathrm{Ca}}(\ref{eq2})} &=&
\left( {{\begin{array}{*{20}c}
 0 \hfill & 0 \hfill & {d_{\mathrm{A}} } \hfill \\
 0 \hfill & 0 \hfill & 0 \hfill \\
 {d_{\mathrm{A}} } \hfill & 0 \hfill & 0 \hfill \\
\end{array} }} \right) \nonumber \\ r_{F_{{2g} \mathrm{A}}(\ref{eq3})} &=&
\left( {{\begin{array}{*{20}c}
 0 \hfill & 0 \hfill & 0 \hfill \\
 0 \hfill & 0 \hfill & {d_{\mathrm{A}} } \hfill \\
 0 \hfill & {d_{\mathrm{A}} } \hfill & 0 \hfill \\
\end{array} }} \right)
\end{eqnarray}
where

\begin{equation}
\label{eq24} d_{\mathrm{A}} = r_{F_{2g} \mathrm{A}(\ref{eq1}),xy}
= \frac{1}{\sqrt {2m_{\mathrm{A}} } }\left( {r_{xy,\mathrm{A}1z} -
r_{xy,\mathrm{A}2z} } \right)
\end{equation}
with the symmetry relations:

\begin{equation}
\label{eq25}
\begin{array}{l}
 r_{xy,\mathrm{A}1z} = - r_{xy,\mathrm{A}2z} = - r_{xz,\mathrm{a}1y} = \\ =
r_{xz,\mathrm{A}2y} = r_{yz,\mathrm{A}1x} =
\\
 = - r_{yz,\mathrm{A}2x} = - r_{yx,\mathrm{A}1z} = r_{yx,\mathrm{A}2z} = \\ = -
r_{zx,\mathrm{A}1y} = r_{zx,\mathrm{A}2y}
\\
 \end{array}
\end{equation}
From these relations:

\begin{equation}
\label{eq26} d_{\mathrm{A}} = r_{F_{2g} \mathrm{A}(\ref{eq1}),xy}
= \sqrt {\frac{2}{m_{\mathrm{A}} }} r_{xy,\mathrm{A}1z}
\end{equation}
The eigenvectors of the F$_{2g}$ modes are linear combinations of
the O- and A-related normalized vectors:

\begin{equation}
\label{eq27} e_{F_{2g} } = c_\mathrm{O} e_{F_{{2g} \mathrm{O}}} +
c_{\mathrm{A}} e_{F_{{2g} \mathrm{A}}}
\end{equation}
Their Raman tensors can be represented as linear combinations of
the O- and A-related Raman tensors

\begin{equation}
\label{eq28} r_{F_{2g} \alpha \beta } = c_\mathrm{O} r_{F_{2g}
\mathrm{O}\alpha \beta } + c_{\mathrm{A}} r_{F_{2g}
\mathrm{A}\alpha \beta }
\end{equation}
The polarized $I_{i\parallel}^{cryst} $ and depolarized $I_{i
\perp }^{cryst} $Raman spectra of the cubic AB$_{'1 / 2}$B$''_{1 /
2}$O$_{3}$ perovskites are determined by the diagonal and
nondiagonal elements of the Raman tensor, respectively:

\begin{equation}
\label{eq29}
\begin{array}{l}
 I_{i\parallel}^{cryst} \sim \left( {\omega _L - \omega _i }
\right)^4\frac{r_{izz}^2 }{\omega _i }\left[ {1 + n\left( {\omega
_i }
\right)} \right] \\
 I_{i \perp }^{cryst} \sim \left( {\omega _L - \omega _i }
\right)^4\frac{r_{ixz}^2 }{\omega _i }\left[ {1 + n\left( {\omega
_i }
\right)} \right] \\
 I_{it}^{cryst} \sim I_{i\parallel}^{cryst} + I_{i \perp}^{cryst} \\
 \end{array}
\end{equation}
From these quantities, the corresponding Raman spectra can be
computed as:

\begin{equation}
\label{eq30}
\begin{array}{l}
 I_{\parallel}^{cryst} \left( \omega \right)\sim \sum\limits_i
{I_{i\parallel }^{cryst} \frac{\Gamma _i }{\left( {\omega - \omega
_i } \right)^2 +
\Gamma _i^2 }} \\
 I_ \perp ^{cryst} \left( \omega \right)\sim \sum\limits_i {I_{i
 \perp
}^{cryst} \frac{\Gamma _i }{\left( {\omega - \omega _i } \right)^2
+ \Gamma
_i^2 }} \\
 I_t^{cryst} \left( \omega \right)\sim \sum\limits_i {I_{it}^{cryst}
\frac{\Gamma _i }{\left( {\omega - \omega _i } \right)^2 + \Gamma _i^2 }} \\
 \end{array}
\end{equation}
The Raman spectra for powders can be found from expressions
(\ref{eq10}).

\begin{table}[!htbp] \caption{Relative values of Raman tensor
components obtained from the first-principles computations of
cubic CAN.}
\begin{tabular}{|c|c|}
\hline  notation & value
\\
\hline a & 0.03084 \\
b & 0.00717 \\
d$_{\mathrm{O}}$ & 0.01572 \\
d$_{\mathrm{Ca}}$ & 0.00375 \\ \hline
\end{tabular}\label{tab1}  \end{table}

\begin{table}[!htbp]
\caption{Eigenvectors for $F_{2g}$ modes.}
\begin{tabular}
{|c|c|c|} \hline coefficients& F$_{2g}$I&
F$_{2g}$II \\
\hline $c_{\mathrm{O}}$& 0.9794&
0.2019 \\
$c_{\mathrm{Ca}}$& -0.2019&
0.9794 \\
\hline
\end{tabular}
\label{tab2}
\end{table}

\begin{table}[!htbp]\begin{ruledtabular}
\caption{Relative intensities of Raman lines obtained in the
first-principles computation for single crystal cubic CAN (T =
300K).}
\begin{tabular}
{c|cccccc} mode & $\nu$ [cm$^{ - 1}$] & $I^{cryst}$ & Geometry &
$I^{powder}_t$ & $ I^{powder}_\parallel$ &
$I^{powder}_\perp$ \\
\hline A$_{1g}$& 910& 8.7388& $\parallel$& 8.7388& 8.7388&
0.0000 \\
E$_{g} $& 660& 2.8332& $\parallel$& 1.6527& 0.9444&
0.7083 \\
F$_{2g}$I& 445& 4.9791& $\perp$& 6.9708& 3.9833&
2.9875 \\
F$_{2g}$II& 101& 11.8411& $\perp$& 16.5776& 9.4729&
7.1047 \\
\end{tabular} \label{tab3}\end{ruledtabular}
\end{table}

\begin{table}[!htbp]
\caption{The relative integral intensities obtained in the
computation of CAN ($P2_1/n$). The stars show the modes, which are
closest to those obtained under the $Fm\overline{3}m$~
constraint.}
\begin{tabular} {|c|ccc|} \hline $\nu$ [cm$^{ - 1}$]&
$I_{i\parallel}^{powder} $& $I_{i \perp }^{powder} $&
$I_{it}^{powder} $ \\
\hline 797*& 1.2034& 0.0013&
1.2048 \\
554*& 0.0114& 0.0086&
0.0200 \\
553*& 0.0149& 0.0100&
0.0249 \\
457*& 0.1927& 0.1442&
0.3368 \\
452*& 0.1698& 0.1273&
0.2971 \\
446*& 0.1443& 0.1082&
0.2526 \\
242*& 0.0306& 0.0228&
0.0534 \\
211*& 0.0169& 0.0127&
0.0296 \\
188*& 0.0324& 0.0241&
0.0565 \\
770& 0.0004& 0.0003&
0.0008 \\
351& 0.0015& 0.0011&
0.0026 \\
732& 0.0027& 0.0020&
0.0047 \\
482& 0.0179& 0.0134&
0.0312 \\
289& 0.0000& 0.0000&
0.0000 \\
331& 0.0168& 0.0113&
0.0281 \\
237& 0.0072& 0.0054&
0.0126 \\
359& 0.0113& 0.0084&
0.0197 \\
564& 0.0030& 0.0022&
0.0052 \\
185& 0.0026& 0.0019&
0.0045 \\
175& 0.0161& 0.0121&
0.0282 \\
136& 0.0109& 0.0077&
0.0186 \\
182& 0.0018& 0.0014&
0.0032 \\
162& 0.0063& 0.0042&
0.0105 \\
339& 0.0079& 0.0025& 0.0104 \\ \hline
\end{tabular}
\label{tab5}
\end{table}

\end{document}